\documentclass[aps,prb,floatfix,superscriptaddress]{revtex4}
\usepackage{ifxetex}
\usepackage{amsmath}
\usepackage{subeqnarray}
\usepackage{cases}
\usepackage{amssymb}
\usepackage{bm}
\usepackage{upgreek}
\usepackage{pxfonts}
\usepackage{extarrows}
\usepackage{upgreek}
\usepackage{float}
\usepackage{graphicx}
\usepackage[english]{babel}
\usepackage{slashed}
\usepackage{cancel}
\usepackage{relsize}
\usepackage[T1]{fontenc}
\usepackage{lmodern}
\usepackage{fancyhdr}
\usepackage{mathrsfs}

\usepackage{dsfont}

\newcommand{\me}{\mathrm{e}}
\newcommand{\mi}{\mathrm{i}}

\newcommand{\dif}{\mathrm{d}}

\allowdisplaybreaks
\begin{document}

\title{Equivalence of Two Approaches to Solving Static Electromagnetic Potentials}
\author{Weiwei Zhao}
\altaffiliation{Current address: Tsung-Dao Lee Institute, Shanghai Jiao Tong University, Shanghai 200240, China}
\affiliation{School of Physics, Southeast University, Nanjing 211189, China}

\author{Hao Jin}
\altaffiliation{School of Physics, Peking University, Beijing 100871, China}
\affiliation{School of Physics, Southeast University, Nanjing 211189, China}

\author{Hao Guo}
\email{guohao.ph@seu.edu.cn}
\affiliation{School of Physics, Southeast University, Nanjing 211189, China}

\begin{abstract}
In electromagnetic statics, the electric scalar potential and magnetic vector potential in a bounded region are typically obtained by solving Poisson's equation with appropriate boundary conditions. Alternatively, these potentials can be computed directly by integrating over the charge or current distributions-extending Coulomb's law and the Biot-Savart law to include both volume and surface contributions. In this paper, we rigorously demonstrate the mathematical equivalence of these two methods-the differential (Poisson's equation) and integral approaches-in regions with uniform material properties. Beyond its theoretical importance, this equivalence offers valuable pedagogical insights, providing students with multiple complementary methods for solving electromagnetic problems. Our explicit examples in both electrostatics and magnetostatics verify the consistency of these approaches and serve as effective teaching tools in advanced electromagnetism courses.
\end{abstract}

\maketitle

\section{Introduction}

The origin of this paper traces back to an electrodynamics midterm exam, where the author, Hao Jin, solved a problem (\textit{Example}.1 in this paper) using the direct integral formulation. This solution led us to investigate whether such an approach could be proven more generally. In response, instructor Hao Guo provided a general proof for electrostatics, and, building upon this foundation, Weiwei Zhao extended the proof to magnetostatics. The general form of these types of problems can be stated as follows:
In electrostatics, Coulomb's law establishes the fundamental relationship between a static charge distribution and the resulting electric potential.  Specifically, the electric potential $\phi(\mathbf{x})$ in free space generated by a charge distribution $\rho(\mathbf{x}')$ is given by:
\begin{align}\label{phis}
\phi(\mathbf{x})=\frac{1}{4\pi\varepsilon}\int\frac{\rho(\mathbf{x}')}{|\mathbf{x}-\mathbf{x}'|}\dif V',
\end{align}
where $\mathbf{x}$ denotes the field point, $\mathbf{x}'$ represents the source point, and $\varepsilon$ is the dielectric constant of the medium. This expression determines the potential up to an arbitrary additive constant, reflecting the gauge freedom inherent in electrostatic potentials.
The connection between this integral formulation and the differential form of electrostatics can be established using the identity $\nabla^2\frac{1}{|\mathbf{x}-\mathbf{x}'|}=-4\pi\delta(\mathbf{x}-\mathbf{x}')$. Through this relationship, one can verify that Eq. (\ref{phis}) satisfies Poisson's equation:
\begin{align}\label{peq}
\nabla^2\phi=-\frac{\rho}{\varepsilon},
\end{align}
which serves as the fundamental governing equation for electrostatic potentials in the presence of charge distributions.

Several important assumptions underlie this result: (1) the entire space is filled with a single homogeneous material (or vacuum) characterized by permittivity $\varepsilon$, and (2) the charge density $\rho$ is well-behaved (i.e., finite and continuous except possibly at isolated points or surfaces). However, realistic physical systems often involve multiple materials with complex geometries and boundaries. In such cases, the standard approach is to solve Poisson's equation (\ref{peq}) in each material domain while enforcing appropriate boundary conditions at interfaces, as discussed comprehensively in standard electrodynamics textbooks.

When boundaries carry electric charge, the surface charge density $\sigma$ must be incorporated into the boundary conditions. Mathematically, this surface charge can be represented as a singular volume charge density using the Dirac-$\delta$-function: $\rho(\mathbf{x})=\sigma(\mathbf{x}')\delta(\mathbf{x}-\mathbf{x}')$, where $\mathbf{x}'$ lies on the boundary surface $S'$. Substituting this expression into Eq. (\ref{phis}) yields the potential due to surface charges:
\begin{align}\label{phisc}
\phi(\mathbf{x})=\frac{1}{4\pi\varepsilon}\int_{\mathbf{x}''\in S'}\dif V'\frac{\sigma(\mathbf{x}'')\delta(\mathbf{x}''-\mathbf{x}')}{|\mathbf{x}-\mathbf{x}''|}
=\frac{1}{4\pi\varepsilon}\iint_{S'}\dif S'\frac{\sigma(\mathbf{x}')}{|\mathbf{x}-\mathbf{x}'|}.
\end{align}
It is crucial to note that this expression is valid only when the materials on both sides of $S'$ are identical. In cases where different media meet at the interface, the potential $\phi$ becomes discontinuous across $S'$, and additional considerations involving surface polarization charges are required.

Equation (\ref{phisc}) represents the direct contribution of surface charges to the electric potential. In the conventional boundary-value approach to solving Poisson's equation, this contribution is implicitly accounted for through the boundary conditions. These two methodologies-the direct integral formulation and the differential equation approach with boundary conditions-are mathematically equivalent, as guaranteed by the uniqueness theorem of electrostatics. Although the surface potential expression (\ref{phisc}) appears in some authoritative texts, primarily in the context of bound surface charges at dielectric interfaces \cite{Griffiths_book, Greiner_book, Ohanian_book, Muller-Kirsten_book}, a rigorous mathematical demonstration of its equivalence to the boundary-value solution of Poisson's equation has not, to our knowledge, been explicitly presented. In fact, while these texts address similar formulas for dielectric media, Eq. (\ref{phisc}) is also applicable to conductor surfaces, and its counterpart in magnetostatics has been rarely discussed. Here, we provide the first rigorous proof that this integral method is equivalent to solving the boundary-value problem for Poisson's equation in both electrostatics and magnetostatics. Exploring this equivalence in detail not only clarifies theoretical concepts but also offers valuable perspectives for students. By comparing these methods, students can gain a deeper understanding of the underlying physics and strengthen their problem-solving skills.

Recent studies in physics education have demonstrated that exposing students to multiple problem-solving strategies-such as the differential and integral approaches in electromagnetism-can significantly enhance their conceptual understanding and analytical capabilities. Our work builds upon these pedagogical insights by providing a rigorous demonstration of the equivalence between these methods, thereby offering educators a powerful resource to enrich advanced electromagnetism curricula.

\section{Proof of Equivalence Between Boundary-Value and Integral Methods}

\subsection{Electrostatic Field Formulation}
For clarity in our discussion of electrostatic field formulation, we distinguish two approaches to determining electrostatic potentials:
1) The \textit{routine method} involving solutions to Poisson's equation with boundary conditions, and
2) The \textit{integral method} that directly sums contributions from all charge distributions.

The routine method requires solving the coupled Poisson equations:
\begin{eqnarray}\label{20}
\left\{\begin{array}{ll}
\nabla^2\phi_1=-\dfrac{\rho}{\varepsilon}, & \mathbf{x} \in \text{region 1} \\
\nabla^2\phi_2=-\dfrac{\rho}{\varepsilon}, & \mathbf{x} \in \text{region 2}
\end{array}\right.
\end{eqnarray}
subject to boundary conditions:
\begin{align}
&\phi_1\big|_{S'}=\phi_2\big|_{S'}, \label{2}\\
&\varepsilon\hat{n}\cdot\left(\nabla\phi_1\big|_{S'}-\nabla\phi_2\big|_{S'}\right)=-\sigma\big|_{S'}. \label{2b}
\end{align}
where $\hat{n}$ is the outward normal vector from region 1 to region 2. The solution typically proceeds as follows: 1) expand $ \phi_{1,2} $ in a set of basis functions that satisfy Laplace's equation; 2) determine the expansion coefficients by applying the boundary conditions; and 3) construct the complete solution via linear superposition.

The integral approach provides a unified expression:
\begin{align}\label{1}
\phi(\mathbf{x})=\frac{1}{4\pi\varepsilon}\left[\iiint_V\frac{\rho(\mathbf{x}')}{|\mathbf{x}-\mathbf{x}'|}\dif V' + \oiint_{S'}\frac{\sigma(\mathbf{x}')}{|\mathbf{x}-\mathbf{x}'|}\dif S'\right].
\end{align}
We demonstrate the equivalence of these methods by verifying that (\ref{1}) satisfies (\ref{20})-(\ref{2b}), with uniqueness guaranteeing this is the sole solution. We present both general and spherical boundary cases for pedagogical clarity.

\begin{figure}[htbp]
\centering
\includegraphics[width=3.5in]{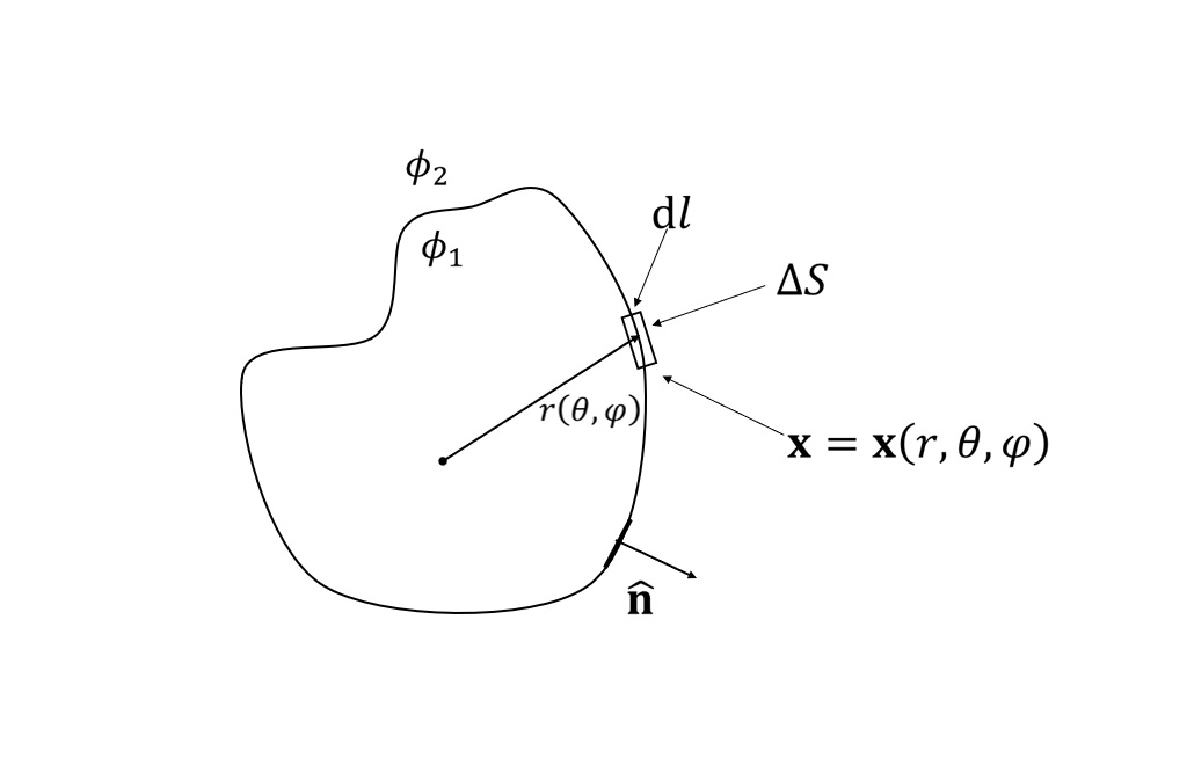}
\caption{Gaussian pillbox construction at surface $S'$ for boundary condition analysis. The pillbox thickness $2\epsilon$ is exaggerated for visualization.}
\label{fig1}
\end{figure}

\subsection{Proof of Equivalence}

We now present a rigorous demonstration of the equivalence between the boundary-value approach and the integral formulation for electrostatic potentials. The proof proceeds in three parts: verification of Poisson's equation, continuity of the potential across boundaries, and satisfaction of the normal derivative boundary condition.


First, we verify that the integral expression in Eq.~(\ref{1}) satisfies Poisson's equation (\ref{20}). Applying the Laplacian operator to the potential yields:
\begin{align}\label{pe1}
\nabla^2\phi(\mathbf{x}) &= \frac{1}{4\pi\varepsilon}\iiint_V \rho(\mathbf{x}')\left(\nabla^2\frac{1}{|\mathbf{x}-\mathbf{x}'|}\right)\dif V'  + \frac{1}{4\pi\varepsilon}\oiint_{S'} \sigma(\mathbf{x}')\left(\nabla^2\frac{1}{|\mathbf{x}-\mathbf{x}'|}\right)\dif S' \notag\\
&= -\frac{\rho(\mathbf{x})}{\varepsilon} - \frac{1}{\varepsilon}\oiint_{S'} \sigma(\mathbf{x}')\delta(\mathbf{x}-\mathbf{x}')\dif S' \notag\\
&= -\frac{\rho(\mathbf{x})}{\varepsilon}
\end{align}
where we have employed the well-known identity $\nabla^2\frac{1}{|\mathbf{x}-\mathbf{x}'|} = -4\pi\delta(\mathbf{x}-\mathbf{x}')$. The surface integral vanishes identically for field points $\mathbf{x}$ not lying on the boundary $S'$, confirming that the potential satisfies Poisson's equation in both regions.


Next, we examine whether the electric potential remains continuous across the boundary.
Although the continuity of $ \phi $ across $ S' $ might seem evident from the form of Eq.~(\ref{phisc}), a careful analysis is required due to the presence of the surface integral term. Consider a point $ \mathbf{x} $ approaching the boundary from either side, with $ r = r' \pm \epsilon $, where $ \epsilon $ is an infinitesimal positive quantity. Expanding the Green's function in Legendre polynomials about the boundary point $ \mathbf{x}' = (r', \theta', \varphi') $ yields:
\begin{align}\label{Lde}
\frac{1}{|\mathbf{x}-\mathbf{x}'|} = \begin{cases}
\sum_{l=0}^\infty \dfrac{(r'-\epsilon)^l}{r'^{l+1}}P_l(\cos\alpha), & \text{for interior approach} \\
\sum_{l=0}^\infty \dfrac{r'^l}{(r'+\epsilon)^{l+1}}P_l(\cos\alpha), & \text{for exterior approach}
\end{cases}
\end{align}
where $ \alpha $ is the angular separation between $ \mathbf{x} $ and $ \mathbf{x}' $. The resulting potential difference across the boundary becomes:
\begin{align}
\phi_1\big|_{S'}-\phi_2\big|_{S'} &\approx \frac{1}{4\pi\varepsilon} \oiint_{S'} \dif S' \, \sigma(\mathbf{x}') \sum_l \left[\left(1 - l\frac{\epsilon}{r'}\right) - \left(1 - (l+1)\frac{\epsilon}{r'}\right)\right] \frac{P_l(\cos\alpha)}{r'} \notag \\
&= \frac{\epsilon}{4\pi\varepsilon} \oiint_{S'} \dif S' \, \sigma(\mathbf{x}') \sum_l \frac{1}{r'^2} P_l(\cos\alpha).
\end{align}
As $ \epsilon \rightarrow 0 $, the continuity condition $ \phi_1\big|_{S'} - \phi_2\big|_{S'} = 0 $ is satisfied, since the leading-order difference vanishes linearly with $ \epsilon $.
However, if the materials on either side of the boundary differ-specifically, if the permittivity is $ \varepsilon_1 $ in region 1 and $ \varepsilon_2 $ in region 2-then the above derivation no longer applies. In other words, the integral approach is valid only when the entire space is occupied by a single, homogeneous medium.

Finally, we demonstrate the validity of the normal derivative boundary condition (\ref{2b}). To this end, we consider a thin Gaussian pillbox enclosing a small patch of $S'$, as illustrated in Fig.~\ref{fig1}. We assume the center of the pillbox is located at $ \mathbf{x} = (r, \theta, \varphi) $, with a thickness of $ 2\epsilon $. The top and bottom surfaces of the pillbox are at $ r_\pm \equiv r \pm \epsilon $, each having an area of $ \Delta S = r^2 \sin\theta \Delta\theta \Delta\varphi $. Compared to the top and bottom surface areas, the side surface area is a higher-order infinitesimal and can therefore be safely neglected.
Multiplying both sides of Eq. (\ref{2b}) by $ \Delta S $ and applying Gauss's theorem, we obtain:
\begin{align}\label{3}
&\left[-\varepsilon\hat{n}\cdot\nabla\phi_2\big|_{S'}+\varepsilon\hat{n}\cdot\nabla\phi_1\big|_{S'}\right]\Delta S
= (\mathbf{D}_2\cdot \hat{n} - \mathbf{D}_1\cdot \hat{n})\Delta S \approx \oiint_{\partial (\text{pillbox})} \mathbf{D} \cdot \dif\mathbf{S} = \iiint_{\text{pillbox}} \nabla \cdot \mathbf{D} \, \dif V \notag\\
= -&\iiint_{\text{pillbox}} \varepsilon \nabla^2 \phi(\mathbf{x}) \, \dif V= -\int_{r_-}^{r_+} \dif l \, \varepsilon \nabla^2 \phi(\mathbf{x}) \, \Delta S = -\int_{r_-(\theta,\varphi)}^{r_+(\theta,\varphi)} \dif l \, \varepsilon \nabla^2 \phi(\mathbf{x}) \, r^2 \sin\theta \, \Delta\theta \Delta\varphi,
\end{align}
where $ \mathbf{D} $ is the electric displacement field, $ \partial (\text{pillbox}) $ denotes the boundary of the pillbox, and the relation $ \dif V = \dif l \Delta S $ is used in the sixth line.
Now substituting Eq. (\ref{1}) into Eq. (\ref{3}), the first term vanishes since $ \rho $ must be zero outside the boundary (as it represents the volume charge density), while the second term gives
\begin{align}\label{4}
&\left(-\varepsilon\hat{n}\cdot\nabla\phi_2\big|_{S'} + \varepsilon\hat{n}\cdot\nabla\phi_1\big|_{S'}\right)\Delta S = -\frac{1}{4\pi} \int_{r'_-(\theta,\varphi)}^{r'_+(\theta,\varphi)} \dif r \oiint_{S'} \dif S' \, \sigma(\mathbf{x}') \left( \nabla^2 \frac{1}{|\mathbf{x} - \mathbf{x}'|} \right) \Delta S \notag\\
=& \int_{r'_-(\theta,\varphi)}^{r'_+(\theta,\varphi)} \dif r \int_0^\pi \dif\theta' \int_0^{2\pi} \dif\varphi' \, r'^2 \sin\theta' \, \sigma(\mathbf{x}') \, \delta(\mathbf{x} - \mathbf{x}') \, \Delta S.
\end{align}
Here, $ \dif l $ is replaced by $ \dif r $ as it represents a radial change, and $ r_\pm $ is changed to $ r'_\pm $ since $ \mathbf{x} $ is evaluated on $ S' $.
Next, using the spherical-coordinate representation of the delta function,
\begin{eqnarray}\label{5}
\delta(\mathbf{x} - \mathbf{x}') = \frac{1}{r'^2 \sin\theta'} \delta(r - r') \delta(\theta - \theta') \delta(\varphi - \varphi'),
\end{eqnarray}
we obtain
\begin{align}\label{6}
 \left(-\varepsilon\hat{n}\cdot\nabla\phi_2\big|_{S'} + \varepsilon\hat{n}\cdot\nabla\phi_1\big|_{S'}\right)\Delta S = & \int_{r'_-(\theta,\varphi)}^{r'_+(\theta,\varphi)} \dif r \int_0^\pi \dif\theta' \int_0^{2\pi} \dif\varphi' \, \sigma(r', \theta', \varphi')  \delta(r - r') \delta(\theta - \theta') \delta(\varphi - \varphi') \Delta S \notag\\
=& \sigma(r(\theta,\varphi), \theta, \varphi) \Delta S,
\end{align}
which finally yields
\begin{eqnarray}\label{7}
\varepsilon\hat{n} \cdot \left( \nabla\phi_1\big|_{S'} - \nabla\phi_2\big|_{S'} \right) = -\sigma\big|_{S'}.
\end{eqnarray}
Hence, the two approaches are indeed equivalent.

From a methodological perspective, the integral approach fundamentally treats surface charge density as a Dirac-$\delta$-singular volume charge density at interfaces, with the interface itself represented as a special volume element of zero thickness. This treatment is particularly advantageous for systems with a uniform permittivity throughout the entire space. By converting boundary conditions into equivalent singular source terms, the formulation provides an alternative, yet equally valid, way to solve electromagnetic problems. This multiplicity of solution methods not only deepens our conceptual understanding of electrodynamics but also offers flexibility in tackling complex scenarios. We will present a direct comparison using a concrete example in the following discussion.
\subsection{Special Case: Spherical Boundary}

If $ S' $ is a geometrically regular boundary, the proof becomes more intuitive. For simplicity, we consider only condition (\ref{2b}). Assume that $ S' $ is a spherical surface of radius $ R_0 $, then the expansion (\ref{Lde}) becomes
\begin{eqnarray}
\dfrac{1}{|\mathbf{x}-\mathbf{x}'|} = \begin{cases} \sum_{l=0}^\infty \dfrac{r^l}{R_0^{l+1}} P_l(\cos\alpha), & r < R_0, \\ \sum_{l=0}^\infty \dfrac{R_0^l}{r^{l+1}} P_l(\cos\alpha), & r > R_0. \end{cases}
\end{eqnarray}
Substituting into Eq. (\ref{1}) and using the addition theorem for spherical harmonics,
\begin{eqnarray}\label{sh}
P_l(\cos\alpha) = \sum_{m=-l}^l \frac{4\pi}{2l+1} Y_{lm}^*(\theta', \varphi') Y_{lm}(\theta, \varphi),
\end{eqnarray}
the left-hand side of Eq. (\ref{2b}) becomes
\begin{align}\label{8}
 \varepsilon \left( \frac{\partial \phi_2}{\partial r} \Big|_{r \rightarrow R_0^+} - \frac{\partial \phi_1}{\partial r} \Big|_{r \rightarrow R_0^-} \right) = & \frac{1}{4\pi} \oiint_{S'} \dif S' \, \sigma(\mathbf{x}') \sum_{l=0}^\infty \left[ (-l-1) \frac{R_0^l}{R_0^{l+2}} - l \frac{R_0^{l-1}}{R_0^{l+1}} \right] P_l(\cos\alpha) \notag\\
=& -\sum_{l=0}^\infty \sum_{m=-l}^l \sigma_{lm} Y_{lm}(\theta, \varphi)= -\sigma(\theta, \varphi),
\end{align}
where $\sigma_{lm} = \int_0^\pi \dif\theta' \sin\theta' \int_0^{2\pi} \dif\varphi' \, \sigma(\theta', \varphi') Y_{lm}^*(\theta', \varphi')$
is the $lm$-th coefficient in the spherical harmonic expansion of $\sigma(\theta, \varphi)$.
This provides an explicit verification of the boundary condition for the spherical case.

\subsection{Magnetostatic Field}

A parallel theorem can be derived in magnetostatics. In the standard approach, the magnetostatic potential is determined by solving Poisson's equation
\begin{equation}\label{PdeA}
\nabla^2\mathbf{A}=-\mu_0\mathbf{J}
\end{equation}
subject to the boundary conditions
\begin{equation}\label{ABD2}
\begin{cases}
(\mathbf{A}_1-\mathbf{A}_2)\big|_{S'}=0,\\
\left(\frac{1}{\mu}\nabla\times\mathbf{A}_1-\frac{1}{\mu}\nabla\times\mathbf{A}_2\right)\big|_{S'}=\hat{n}\times\vec{\alpha},
\end{cases}
\end{equation}
where $\vec{\alpha}$ represents the surface current density. In a similar fashion, the magnetic vector potential can also be obtained by generalizing the Biot-Savart law:
\begin{align}\label{AVS1}
\mathbf{A}(\mathbf{x})=\frac{\mu}{4\pi}\int\dif V'\frac{\mathbf{J}(\mathbf{x}')}{|\mathbf{x}-\mathbf{x}'|}+\frac{\mu}{4\pi}\oiint_{S'}\dif S'\frac{\vec{\alpha}(\mathbf{x}')}{|\mathbf{x}-\mathbf{x}'|},
\end{align}
where the second integral is implied by substituting $\mathbf{J}(\mathbf{x})=\vec{\alpha}(\mathbf{x}')\delta(\mathbf{x}-\mathbf{x}')$, analogous to the evaluation in Eq. (\ref{phisc}).
The uniqueness theorem ensures that this is the solution to the boundary value problem specified by Eqs. (\ref{PdeA}) and (\ref{ABD2}). We again emphasize that this theorem applies only in situations where a single material type is present.

\begin{figure}[htbp]
\centering  \includegraphics[width=2.8in]{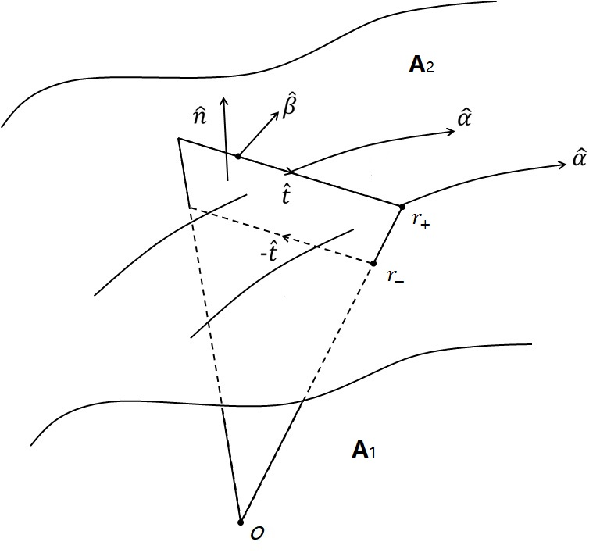}
\caption{A small rectangular loop on the surface $S'$.}
\label{fig2}
\end{figure}

\textit{Proof}:
As with Eq. (\ref{pe1}), Eq. (\ref{AVS1}) is the solution to Poisson's equation (\ref{PdeA}). For boundary conditions, we consider an Amperian loop as shown in Fig. \ref{fig2}, with length $\Delta l$ and width $2\epsilon$, where $\Delta l \gg \epsilon$ (For endpoints inside/outside the boundary, the distances from the origin are denoted by $r'_-$ and $r'_+$, with $r'_+ - r'_- = 2\epsilon$). Let $\hat{t}$ denote the tangent direction along the length, and $\hat{n}$ the normal direction of the boundary.
Define $\hat{\beta} = \hat{n} \times \hat{t}$, indicating the normal direction of the Amperian rectangle.
Assuming $\mathbf{J}$ is stationary, then $\mathbf{A}$ satisfies the Coulomb gauge condition $\nabla \cdot \mathbf{A} = 0$.
By applying Stokes's theorem to evaluate the contour integral of $\mathbf{H}$ along the Amperian rectangle, we obtain
\begin{align}\label{CIofA}
\oint\mathbf{H}\cdot\dif \mathbf{l} &= \oint\frac{1}{\mu}\nabla\times\mathbf{A}\cdot\dif \mathbf{l}= \oiint_{\Delta S}\frac{1}{\mu}\nabla\times(\nabla\times\mathbf{A})\cdot\dif \mathbf{S}= -\oiint_{\Delta S}\frac{1}{\mu}\nabla^2\mathbf{A}\cdot\dif \mathbf{S}\notag \\
&= -\oiint_{\Delta S}\frac{1}{\mu}\frac{\mu}{4\pi}\oiint_{S'}\dif S'(-4\pi)\delta(\mathbf{x}-\mathbf{x}')\vec{\alpha}(\mathbf{x}')\cdot\dif \mathbf{S}= \iint_{\Delta S}\dif S\oiint_{S'}\vec{\alpha}(\mathbf{x}')\cdot \hat{\beta}\delta(\mathbf{x}-\mathbf{x}')\dif S'.
\end{align}
Here, the expression in Eq. (\ref{AVS1}) has been used. The area $\Delta S$ is the region surrounded by the Amperian loop, which is approximately given by
\begin{align}\label{DS}
\Delta S \approx 2\epsilon \Delta l = \iint_{\Delta S}\dif S \approx \Delta l \int_{r'_-}^{r'_+}\dif r.
\end{align}
Moreover, $\vec{\alpha} \cdot \dif \mathbf{S} = \vec{\alpha} \cdot \hat{\beta} \dif S$ because $\hat{\beta}$ is the normal vector of the infinitesimal patch $\dif \mathbf{S}$.
Similarly to the electric case, the boundary $S'$ is parameterized by $r'(\theta', \phi')$, so that $\dif S' = r'^2 \sin \theta' \dif \theta' \dif \varphi'$. Using Eqs. (\ref{DS}) and (\ref{5}), Eq. (\ref{CIofA}) simplifies to
\begin{align}\label{CIofA0}
\oint\mathbf{H}\cdot\dif \mathbf{l} &\approx \int_{r'_-}^{r'_+}\dif r\int_0^{\pi}\dif \theta'\int_0^{2\pi}\dif \varphi'\vec{\alpha}\cdot \hat{\beta}\delta(r-r')\delta(\theta-\theta')\delta(\varphi-\varphi')\Delta l\notag \\
&= \vec{\alpha}(r(\theta,\varphi),\theta,\varphi)\cdot \hat{\beta}\Delta l.
\end{align}
Here, the coordinate set $(r,\theta,\varphi)$ denotes a point in $\Delta S$, and the delta functions enforce that it lies at the boundary $S'$, at the intersection of $\Delta S$ and $S'$. Next, applying $\hat{\beta} = \hat{n} \times \hat{t}$, we find
\begin{align}\label{CIofA1}
\oint\mathbf{H}\cdot\dif \mathbf{l} \approx \vec{\alpha}\cdot\hat{\beta}\Delta l = \vec{\alpha}\cdot(\hat{n}\times\hat{t})\Delta l = \hat{t}\cdot(\vec{\alpha}\times\hat{n})\Delta l.
\end{align}
Note that the contour integral of $\mathbf{H}$ can also be evaluated as
\begin{align}\label{CIofA2}
\oint\mathbf{H}\cdot\dif \mathbf{l} \approx \left(\mathbf{H}_2-\mathbf{H}_1\right)\big|_{S'}\cdot\hat{t}\Delta l = \frac{1}{\mu}\left(\nabla\times\mathbf{A}_2-\nabla\times\mathbf{A}_1\right)\Big|_{S'}\cdot\hat{t}\Delta l,
\end{align}
and the arbitrariness of $\hat{t}$ leads to
\begin{align}\label{CIofA3}
\left(\frac{1}{\mu}\nabla\times\mathbf{A}_2-\frac{1}{\mu}\nabla\times\mathbf{A}_1\right)\Bigg|_{S'} = \vec{\alpha}\times\hat{n}.
\end{align}
Thus, the solution in Eq. (\ref{AVS1}) satisfies the boundary condition. The continuity of $\mathbf{A}$ across the surface can be verified in the same way as that for $\phi$, and we omit the details here.

\section{Illustrative Examples}

To demonstrate the practical equivalence between the boundary-value and integral approaches, we examine two characteristic problems in electrostatics and magnetostatics.

\textit{Example 1}: Consider a thin dielectric sphere of radius $ R_0 $, on which a surface charge density $ \sigma = \sigma_0 \cos \theta $ is maintained by some external mechanism. Determine the electric potential both inside and outside the sphere.

\textit{Routine Method}: The electric potentials inside and outside the sphere can be expressed as $\phi_1(r,\theta) = \sum_{l=0}^{\infty} A_l r^l P_l(\cos \theta)$ for $r<R_0$, and $\phi_2(r,\theta) = \sum_{l=0}^{\infty} \frac{B_l}{r^{l+1}} P_l(\cos \theta)$ for $r>R_0$.
Applying the boundary conditions, we get $ A_l R_0^l = \frac{B_l}{R_0^{l+1}}$ and $\sum_{l=0}^{\infty} \left( l A_l R_0^{l-1} + (l+1) \frac{B_l}{R_0^{l+2}} \right) P_l(\cos \theta) = \frac{\sigma}{\varepsilon_0}$.
Since $ \sigma = \sigma_0 P_1(\cos \theta) $, we have
\begin{eqnarray} \label{E72}
A_1 + \frac{2 B_1}{R_0^3} = \frac{\sigma_0}{\varepsilon_0}, \quad l A_l R_0^{l-1} + (l+1) \frac{B_l}{R_0^{l+2}} = 0 \quad \textrm{for} \quad l \geq 3.
\end{eqnarray}
Finally, the solutions for the potentials are
\begin{eqnarray}
\begin{cases}
\phi_1(r,\theta) = \frac{\sigma_0}{3 \varepsilon_0} r \cos \theta, & r < R_0, \\
\phi_2(r,\theta) = \frac{R_0^3 \sigma_0}{3 r^2 \varepsilon_0} \cos \theta, & r > R_0.
\end{cases}
\end{eqnarray}

\textit{Integral Method}:
 Using Eq. (\ref{1}), the electric potential is given by
\begin{align}
\phi(\mathbf{x}) \equiv \phi(r,\theta) = \frac{1}{4 \pi \varepsilon_0} \oiint \dif S' \frac{\sigma(\mathbf{x}')}{|\mathbf{x} - \mathbf{x}'|},
\end{align}
where $ \mathbf{x}' = (R_0, \theta', \varphi') $ denotes a point on the sphere, and $ \sigma(\mathbf{x}') = \sigma(\theta') = \sigma_0 \cos \theta' $. Let the angle between $ \mathbf{x} $ and $ \mathbf{x}' $ be $ \alpha $. Using the expression from Eq. (\ref{sh}), the electric potential inside the sphere is
\begin{widetext}
\begin{align}
\phi_1(r,\theta) &= \frac{1}{4 \pi \varepsilon_0} \oiint \dif S' \sum_{l=0}^{+\infty} \sum_{m=-l}^l \frac{r^l}{R_0^{l+1}} \sigma(\theta') P_l(\cos \alpha) \notag \\
&= \sum_{l=0}^{+\infty} \sum_{m=-l}^l \frac{r^l R_0^2}{4 \pi \varepsilon_0 R_0^{l+1}} \int_0^{2\pi} \dif \varphi' \int_0^\pi \dif \theta' \sin \theta' \sigma(\theta') \frac{4\pi}{2l+1} Y_{lm}^*(\theta', \varphi') Y_{lm}(\theta, \varphi).
\end{align}
\end{widetext}
Next, substitute the expression for $ Y_{lm}^*(\theta',\varphi') $:
\begin{align}
Y_{lm}^*(\theta',\varphi') = \sqrt{\frac{2l+1}{4\pi} \frac{(l-m)!}{(l+m)!}} P_l^m(\cos \theta') \me^{-\mi m\varphi'},
\end{align}
and evaluate the integral over $ \varphi' $. Only the $ m = 0 $ term contributes, yielding
\begin{align}
\phi_1(r,\theta) = \sum_{l=0}^{+\infty} \frac{r^l}{2 R_0^{l-1} \varepsilon_0} \left[ \int_0^\pi \dif \theta' \sin \theta' \sigma(\theta') P_l(\cos \theta') \right] P_l(\cos \theta) = \frac{\sigma_0}{3 \varepsilon_0} r \cos \theta,
\end{align}
where we used the orthogonality relation $\int_{-1}^1 \dif x P_l(x) P_{l'}(x) = \frac{2}{2l+1} \delta_{ll'}$.
Similarly, the electric potential outside the sphere is
\begin{align}
\phi_2(r,\theta) = \frac{R_0^3 \sigma_0}{3 r^2 \varepsilon_0} \cos \theta.
\end{align}
This result agrees with that obtained by the routine method. The main advantage of this approach is that the electric potential both inside and outside the interface can be computed using a unified technique (i.e., evaluating the same integral).

To explicitly demonstrate the equivalence theorem in magnetostatics, we choose an example from Griffiths's book \cite{Griffiths_book}.

\textit{Example}.2: A spherical thin shell of radius $ R_0 $, carrying a uniform surface charge $ \sigma $, is set spinning at an angular velocity $ \vec{\omega} = \omega \hat{e}_z $. Find the magnetic field $ \mathbf{H} $ inside and outside the sphere.

\textit{Routine Method}: We can first solve Eq. (\ref{PdeA}) for $ \mathbf{A} $, subject to the boundary conditions (\ref{ABD2}), and then obtain the magnetic field via $ \mathbf{H} = \frac{1}{\mu_0} \nabla \times \mathbf{A} $. However, this process is tedious, requiring at least six sets of coefficients to specify $ \mathbf{A} $ both inside and outside the spherical shell, since $ \mathbf{A} $ is a vector field. A simpler and equivalent method is to introduce the magnetic scalar potential $ \phi_m $ \cite{GSH_book,ZhaoYM_book}, which satisfies Laplace's equation $ \nabla^2 \phi_m = 0 $ inside and outside the sphere, as the electric current is only distributed on the spherical shell.
The boundary conditions are given by:
\begin{align}
B_{1r} |_{r=R_0} = B_{2r} |_{r=R_0}, \quad \hat{e}_r \times (\mathbf{H}_2 - \mathbf{H}_1) |_{r=R_0} = \vec{\alpha},
\end{align}
where $ B_{1,2r} $ is the radial component of $ \mathbf{B}_{1,2} $, and $ \vec{\alpha}(\mathbf{x}) = \sigma \mathbf{v} = \sigma \vec{\omega} \times \mathbf{x} $ represents the surface electric current.
Note that $ \mathbf{H} = -\nabla \phi_m $, so the boundary conditions for the magnetic scalar potential are:
\begin{align}
\frac{\partial \phi_{m1}}{\partial r} \Big|_{r=R_0} = \frac{\partial \phi_{m2}}{\partial r} \Big|_{r=R_0}, \quad
\left( - \frac{1}{R_0} \frac{\partial \phi_{m2}}{\partial \theta} + \frac{1}{R_0} \frac{\partial \phi_{m1}}{\partial \theta} \right) \Big|_{r=R_0} = \sigma \omega R_0 \sin\theta.
\end{align}

The general solution can be expressed as:
\begin{align}
\begin{cases}
\phi_{m1}(r, \theta) = \sum_{l} A_l r^l P_l(\cos\theta), & r < R_0, \\
\phi_{m2}(r, \theta) = \sum_{l} \frac{B_l}{r^{l+1}} P_l(\cos\theta), & r > R_0.
\end{cases}
\end{align}
Substituting these into the boundary conditions, we get $\frac{B_1}{R_0^2} - A_1 R_0 = \sigma \omega R_0^2$, and $A_l = -\frac{(l+1) B_l}{l R_0^{2l+1}}$, $\frac{B_l}{R_0^{l+1}} = A_l R_0^l$ for $l\neq 1$. Hence, the magnetic scalar potential is:
\begin{align}
\begin{cases}
\phi_{m1} = - \frac{2 R_0 \sigma}{3} \vec{\omega} \cdot \mathbf{x}, & r < R_0, \\
\phi_{m2} = \frac{R_0^4 \sigma}{3 r^3} \vec{\omega} \cdot \mathbf{x}, & r > R_0.
\end{cases}
\end{align}
The magnetic field is then:
\begin{align}\label{H1}
\mathbf{H} = -\nabla \phi_m = \left\{
\begin{array}{cc}
\frac{2 R_0 \sigma}{3} \vec{\omega}, & \text{if } r < R_0, \\
\frac{R_0^4 \sigma}{3} \left[ \frac{3 (\vec{\omega} \cdot \mathbf{x}) \mathbf{x}}{r^5} - \frac{\vec{\omega}}{r^3} \right], & \text{if } r > R_0.
\end{array}
\right.
\end{align}

\textit{Integral Method}:
Let the coordinates of a field point be $ \mathbf{x} = (r, \theta, \varphi) $, and the coordinates of a point on the spherical shell be $ \mathbf{x}' = (R_0, \theta', \varphi') $.
The magnetic vector potential is given by:
\begin{align}
\mathbf{A}(\mathbf{x}) = \frac{\mu_0}{4\pi} \oiint \frac{\vec{\alpha}(\mathbf{x}')}{|\mathbf{x} - \mathbf{x}'|} \, \dif S'= \frac{\mu_0 R_0^3 \sigma \omega}{4\pi} \int_0^\pi \dif \theta' \int_0^{2\pi} \dif \varphi' \frac{\sin^2 \theta'}{|\mathbf{x} - \mathbf{x}'|} \hat{e}_{\varphi'}.
\end{align}
Note that this integral requires some technique, because the vector $\hat{e}_{\varphi'}$ depends on $\mathbf{x}'$, and therefore also participates in the integration. To facilitate the calculation, we need to express it as a linear combination of the basis vectors:
\begin{align}\label{EEt3}
\hat{e}_{\varphi'} = -\sin\varphi' \hat{e}_x + \cos\varphi' \hat{e}_y = \frac{1}{2} \left[ (\mi \hat{e}_x + \hat{e}_y) \me^{\mi \varphi'} - (\mi \hat{e}_x - \hat{e}_y) \me^{-\mi \varphi'} \right].
\end{align}
Expanding $ \frac{1}{|\mathbf{x} - \mathbf{x}'|} $ in terms of Legendre polynomials, we get:
\begin{align}\label{H2}
\mathbf{A}(\mathbf{x}) = \left\{
\begin{array}{ll}
\frac{\mu_0 R_0^3 \sigma \omega}{4\pi} \sum_{l=0}^{\infty} \frac{r^l}{R_0^{l+1}} \int_0^\pi \dif \theta' \int_0^{2\pi}\dif \varphi' \sin^2 \theta' P_l(\cos \alpha) \hat{e}_{\varphi'}, & \text{if } r < R_0, \\
\frac{\mu_0 R_0^3 \sigma \omega}{8\pi} \sum_{l=0}^{\infty} \frac{R_0^l}{r^{l+1}} \int_0^\pi \dif \theta' \int_0^{2\pi}\dif \varphi' \sin^2 \theta' P_l(\cos \alpha) \hat{e}_{\varphi'}, & \text{if } r > R_0,
\end{array}
\right.
\end{align}
where $ \alpha $ is the angle between $ \mathbf{x} $ and $ \mathbf{x}' $.
Next, by expanding $ P_l(\cos\alpha) $ in terms of spherical harmonics and using Eq. (\ref{EEt3}), the integral over $\varphi'$ in Eq. (\ref{H2}) can be evaluated as follows:
\begin{align}\label{EEt4}
 \int_0^{2\pi} \dif \varphi'  P_l(\cos\alpha) \hat{e}_{\varphi'}=& \pi   \left[ \frac{\mi \hat{e}_x + \hat{e}_y}{l(l+1)} P_l^1(\cos\theta') P_l^1(\cos\theta) \me^{\mi \varphi} \right.
-\left.  l(l+1)(\mi \hat{e}_x - \hat{e}_y) P_l^{-1}(\cos\theta') P_l^{-1}(\cos\theta) \me^{-\mi \varphi} \right] \notag \\
=&  \frac{2\pi }{l(l+1)} P_l^1(\cos\theta') P_l^1(\cos\theta) \hat{e}_\varphi,
\end{align}
where we have used the relation $\int_0^{2\pi} \dif \varphi' \me^{\mi m \varphi'} = 2\pi \delta_{m0}$ in the second line, and the relation
$
P_l^{-1}(\cos\theta) = -\frac{1}{l(l+1)} P_l^1(\cos\theta)
$
in the last line. Additionally, by utilizing the identity
$
P_l^1(\cos\theta') = -\sin\theta' P_l'^1(\cos\theta'),
$
we find
$
\int_0^\pi \dif \theta' \sin^2\theta' P_l^1(\cos\theta') = -\frac{4}{3} \delta_{l1}.
$
Substituting this result into Eq. (\ref{EEt4}), we obtain
\begin{align}\label{Et5}
\quad \int_0^{\pi} \dif \theta' \int_0^{2\pi} \dif \varphi' \sin^2\theta' P_l(\cos\alpha) \hat{e}_{\varphi'} = \frac{8\pi}{3} \delta_{l1} \sin\theta \hat{e}_\varphi.
\end{align}
Finally, we obtain the vector potential $\mathbf{A}(\mathbf{x})$ as
\begin{align}
\mathbf{A}(\mathbf{x}) = \left\{
\begin{array}{ll}
\frac{\mu_0 R_0 \sigma \omega}{3} r \sin\theta \hat{e}_\varphi = \frac{\mu_0 R_0 \sigma}{3} \vec{\omega} \times \mathbf{x}, & \text{if } r < R_0, \\
\frac{\mu_0 R_0^4 \sigma \omega}{3} \frac{1}{r^2} \sin\theta \hat{e}_\varphi = \frac{\mu_0 R_0^4 \sigma}{3r^3} \vec{\omega} \times \mathbf{x}, & \text{if } r > R_0.
\end{array}
\right.
\end{align}
The corresponding magnetic field is then
\begin{align}
\mathbf{H} = \frac{1}{\mu_0} \nabla \times \mathbf{A} = \left\{
\begin{array}{ll}
\frac{2R_0 \sigma}{3} \vec{\omega}, & \text{if } r < R_0, \\
\frac{R_0^4 \sigma}{3} \left[\frac{3 (\vec{\omega} \cdot \mathbf{x}) \mathbf{x}}{r^5} - \frac{\vec{\omega}}{r^3}\right], & \text{if } r > R_0.
\end{array}
\right.
\end{align}
This result is in perfect agreement with Eq. (\ref{H1}).

\section{Summary and Implications}
In this paper, we have demonstrated that the integral expressions for static electromagnetic potentials-generalizations of Coulomb's law and the Biot-Savart law-are mathematically equivalent to solving Poisson's equation with appropriate boundary conditions in uniform media. Our explicit examples in both electrostatics and magnetostatics verify the practical consistency of these approaches and provide alternative perspectives that are valuable for problem-solving.

From an educational standpoint, this equivalence has significant implications for teaching advanced electromagnetism. By presenting both the integral and differential approaches side by side, educators can offer students multiple tools to tackle complex problems, fostering a deeper understanding of the interconnectedness of physical laws. This comparative approach reinforces key concepts such as gauge invariance and the uniqueness theorem while encouraging students to critically analyze the role of boundary conditions and singularities. Moreover, by bridging theoretical formulations with practical computational techniques, our work provides a versatile resource that can be readily integrated into undergraduate and graduate curricula.

Ultimately, this paper not only reinforces core concepts in electromagnetic theory but also serves as an effective teaching tool, promoting active learning and critical thinking. We hope that our rigorous demonstration and the accompanying pedagogical insights will inspire further research on integrating multiple solution strategies into the teaching of electromagnetism.

H. Guo acknowledges support from the Teaching Research Foundation of Southeast University (Grant No. 5207022106A) and the National Natural Science Foundation of China (Grant No. 12074064).


\begin{thebibliography}{6}
\expandafter\ifx\csname natexlab\endcsname\relax\def\natexlab#1{#1}\fi
\expandafter\ifx\csname bibnamefont\endcsname\relax
  \def\bibnamefont#1{#1}\fi
\expandafter\ifx\csname bibfnamefont\endcsname\relax
  \def\bibfnamefont#1{#1}\fi
\expandafter\ifx\csname citenamefont\endcsname\relax
  \def\citenamefont#1{#1}\fi
\expandafter\ifx\csname url\endcsname\relax
  \def\url#1{\texttt{#1}}\fi
\expandafter\ifx\csname urlprefix\endcsname\relax\def\urlprefix{URL }\fi
\providecommand{\bibinfo}[2]{#2}
\providecommand{\eprint}[2][]{\url{#2}}

\bibitem[{\citenamefont{Griffiths}(1999)}]{Griffiths_book}
\bibinfo{author}{\bibfnamefont{D.~J.} \bibnamefont{Griffiths}},
  \emph{\bibinfo{title}{Introduction to Electrodynamics}}
  (\bibinfo{publisher}{Prentice Hall}, \bibinfo{address}{New Jersey},
  \bibinfo{year}{1999}).

\bibitem[{\citenamefont{Greiner}(1998)}]{Greiner_book}
\bibinfo{author}{\bibfnamefont{W.}~\bibnamefont{Greiner}},
  \emph{\bibinfo{title}{Classical Electrodynamics}}
  (\bibinfo{publisher}{Springer}, \bibinfo{year}{1998}).

\bibitem[{\citenamefont{Ohanian}(1988)}]{Ohanian_book}
\bibinfo{author}{\bibfnamefont{H.~C.} \bibnamefont{Ohanian}},
  \emph{\bibinfo{title}{Classical Electrodynamics}}
  (\bibinfo{publisher}{Prentice Hall}, \bibinfo{year}{1988}).

\bibitem[{\citenamefont{Muller-Kirsten}(2004)}]{Muller-Kirsten_book}
\bibinfo{author}{\bibfnamefont{H.~J.~W.} \bibnamefont{Muller-Kirsten}},
  \emph{\bibinfo{title}{Electrodynamics: an introduction, including quantum
  effects}} (\bibinfo{publisher}{World Scientific},
  \bibinfo{address}{Singapore}, \bibinfo{year}{2004}).

\bibitem[{\citenamefont{Guo}(2008)}]{GSH_book}
\bibinfo{author}{\bibfnamefont{S.~H.} \bibnamefont{Guo}},
  \emph{\bibinfo{title}{Electrodynamics}} (\bibinfo{publisher}{Higher Education
  Press}, \bibinfo{address}{Beijing}, \bibinfo{year}{2008}).

\bibitem[{\citenamefont{Zhao}(2016)}]{ZhaoYM_book}
\bibinfo{author}{\bibfnamefont{Y.~M.} \bibnamefont{Zhao}},
  \emph{\bibinfo{title}{Electrodynamics}} (\bibinfo{publisher}{Science Press},
  \bibinfo{address}{Beijing}, \bibinfo{year}{2016}).

\end{thebibliography}

\end{document}